# Critical Clearing Time Sensitivity for Differential-Algebraic Power System Model

Chetan Mishra, *Member, IEEE*

*Abstract*—Standard power systems are modeled using differential-algebraic equations (DAE). Following a transient event, voltage collapse can occur as a bifurcation of the transient load flow solutions which is marked by the system trajectory reaching a singular surface in state space where the voltage causality is lost. If the system is under such a risk, preventive control decisions such as changes in AVR setpoints need to be taken in order to enhance the stability. In this regard, the knowledge of sensitivity of critical clearing time (CCT) to controllable system parameters can be of great help. The stability boundary of DAE systems is more complicated than ODE systems where in addition to stable manifold of unstable equilibrium points (UEP) and periodic orbits, singular surface plays an important role. In the present work, we derive the expressions for CCT sensitivity for a generic DAE model using trajectory sensitivities with applications to power system transient stability analysis (TSA) and preventive control. The results are illustrated for multiple test systems which are then validated against computationally intensive time domain simulations (TDS).

## I. Introduction

Due to load modeling challenges such as lack of rich data for validation and unavailability of universally accepted dynamic load models, a majority of the utilities still use static load models in at least some parts of their systems which leads to the overall system being modeled as a set of DAEs. It is well known that DAE models can have certain regions in state space called singular surfaces which are characterized by singularity of algebraic equation(s). The dynamics on those parts of state space cannot be studied using these models [1] because the algebraic states like load bus voltages lose causal relationships with dynamic states like generator rotor angle, speed, etc. A type of local bifurcation of equilibria that is a characteristic of such models is singularity induced bifurcation [2] where one or more equilibrium points (EP) merge with the singular surface. Trajectories reaching a singular surface have been shown to have a strong tendency to suffer from a voltage collapse [3].

Lately, utilities all over the world are having voltage stability concerns owing to the retirements of conventional generators which results in loss of voltage controllability. Furthermore, utilities tend to maximize the utilization of the existing transmission network in certain regions owing to the difficulties in building new right-of ways to supply the increasing demand which makes matters worse. In the past, voltage collapse was studied as a small signal problem [4] resulting from the saddle node bifurcation (SNB) of load flow solutions where a stable equilibrium point (SEP) merges with a UEP on its stability boundary and vanishes. However, it was shown that during transient conditions, voltage collapse can occur in a different manner [5]. Trajectories passing through the singular surface may bifurcate and settle to an infeasible (low voltage) point. In [6], Hiskens and Hill showed that operation in the vicinity of a singular surface or trajectories intersecting it was associated with sudden reductions in voltage or voltage instability. Therefore, there has been a great focus in the past on analyzing the stability of DAE systems with the purpose of incorporating voltage stability into the traditional TSA [7].

TSA is concerned with estimating CCT, which refers to the maximum time that can be taken to clear a fault while remaining stable. Since CCT is a function of system conditions, a knowledge of its dependence on various system parameters could be fairly helpful when figuring out effective preventive control decisions to enhance CCT for critical faults and/or being able to quickly analyze a range of operation conditions using sensitivities. In this regard, Ayasun [8] reduced the multimachine system to single machine infinite bus system to evaluate CCT sensitivities which is computationally efficient yet approximate. Nguyen [9] and Laufenberg [10] computed sensitivity of angle and speed trajectory in the post fault phase w.r.t fault clearing time which are expected to grow for marginally stable trajectories. Nguyen also computed CCT sensitivities by approximating the relevant portion of stability boundary by constant energy surface passing through the controlling unstable equilibrium point (CUEP). One of the more recent works by Dobson et.al. [11] does not make this approximation for stability boundary and simply uses a local characterization of it to give more accurate estimates. His derivation is for unconstrained ODE systems and an extension is proposed for DAE systems under the assumption that no portion of the singular surface lies on the stability boundary of the SEP of interest for the range of parameter values under study. Our recent work [12] deals with deriving the same for systems with constraints arising from protection devices and various other limits. To summarize, none of the previous works derive the CCT sensitivity for faults becoming unstable due to singularity (voltage collapse) which is the focus of this work.

In Section II. , the stability theory for DAE systems with emphasis on the role of singular surface is briefly discussed. The main contribution of this paper which is the derivation of expressions for CCT sensitivity for different instability phenomena is presented in Section III. The numerical as well as computational aspects of the overall process are discussed in

Chetan Mishra (email: chetan.mishra@dominionenergy.com) is with Dominion Energy, Richmond, VA-23220, USA.

Section IV. Finally, the derived expressions are validated against TDS on a few interesting low dimensional systems along with visual insights into the qualitative changes in the stability boundary with parameter variations in Section V.

## II. System Model And Stability Theory of DAE Systems

### A. DAE System Dynamics

A generic power system DAE model is of the form,
$$\dot{x} = f(x,y) \qquad (1)$$
$$0 = g(x,y)$$

Here, $x \in R^n$ are dynamic states such as generator rotor angles, generator flux linkages, etc. and $y \in R^m$ are algebraic states such as load bus voltages and phase angles making the overall state space as $R^{m+n}$. The system evolves on a lower dimensional constraint set $\Gamma$ (largely $n$ dimensional) given by,
$$\Gamma = \{(x,y) \in R^{m+n} | g(x,y) = 0\} \qquad (2)$$

A point $(\bar{x}, \bar{y}) \in \Gamma$ is an equilibrium point if $f(\bar{x}, \bar{y}) = 0$. As the system undergoes discrete changes (line tripping, etc.), $\Gamma$ undergoes discrete changes with the system trajectory jumping to the new constraint set. From equation (1), $x$ coordinate of the trajectory varies smoothly in time unlike the $y$ coordinate.

At the points in $\Gamma$ where $\frac{\partial g}{\partial y}$ is invertible (referred to as regular points), by implicit function theorem [13], $y$ can locally be written as a function of $x$. This enables reducing system (1) locally to an ODE system which guarantees the existence and uniqueness of solutions of the DAE system on the regular points. The surface of points where the invertibility condition is not met and therefore the trajectories do not exist is called the singular surface which is denoted by $S$ and is defined as,
$$S = \{x,y \in \Gamma \,|\, \Delta(x,y) = \det\left(\frac{\partial g}{\partial y}\right) = 0\} \qquad (3)$$

Overall, $\Gamma$ is comprised of multiple disjoint subsets of regular points (typically $n$ dimensional submanifolds of the ambient space $R^{n+m}$ [14]) on which the dynamics exist and which are separated by components of $S$.

### B. Characterization of Quasi Stability Boundary

Since the trajectories cannot cross $S$, our region of interest for TSA is one such subset of regular points denoted by $\Gamma^s$ which contains the SEP $(x^s, y^s)$ of interest along with its stability region $A(x^s, y^s)$. Here, $A(x^s, y^s)$ is defined as,
$$A(x^s, y^s) = \{(x^0, y^0) \in \Gamma^s | \lim_{t \to \infty}(\varphi^x((x^0,y^0),t) \qquad (4)$$
$$\to x^s, \varphi^y((x^0, y^0), t) \to y^s)\}$$

Where, $(x(t), y(t)) = (\varphi^x((x^0,y^0),t), \varphi^y((x^0,y^0),t))$ is the solution to (1) for the boundary condition $(x(0), y(0)) = (x^0, y^0)$ with $\varphi$ denoting the flow of vector field in (1). For rest of the paper, $S$ will be used to represent specific component(s) of the singular surface which separate $\Gamma^s$ from other subsets of $\Gamma$. Venkatasubramanian et.al [15] presented a completed characterization of the stability boundary for DAE systems. In this section, we will briefly discuss the important parts of their result which will be helpful in our derivation. Broadly speaking, the stability boundary $(\partial A(x^s, y^s))$ is comprised of some components of $S$, surface on which the trajectories converge to $S$ along and stable manifolds of some unstable equilibrium points (UEP). Instead of the stability boundary which can have a very complex structure, the focus is on characterization of quasi stability boundary which is the boundary of enclosure of $A(x^s, y^s)$ i.e. $\partial \bar{A}(x^s, y^s))$ and is more relevant from engineering point of view.

To help analyze DAE systems using the existing tools for ODE systems, a regularized version of the system was proposed in [15] as shown below.
$$\dot{x} = \Delta(x,y) \times f(x,y) \qquad (5)$$
$$\dot{y} = \kappa(x,y) = -adj\left(\frac{\partial g(x,y)}{\partial y}\right) \times \frac{\partial g(x,y)}{\partial x} \times f(x,y)$$

Without the loss of generality, it is also assumed that $\Delta(x,y) > 0$ inside $\Gamma^s$. Consequently, inside $\Gamma^s$, the above system is equivalent to the original system in (1) which is why their invariant sets including $A(x^s, y^s)$ are identical. The most attractive quality of the above system is that it no longer has singularity problems meaning the dynamics are globally defined. This enables the analysis of dynamics of (1) in the vicinity of $S$ using (5) which is otherwise difficult.

Moving on, there are two important categories of points in $S$. The first category is called semi-singular points at which the transformed system's (Eqn. (5)) trajectory is tangential to $S$ (boundary between shaded and unshaded regions) as shown in Figure 1. These are defined as,
$$\Xi = \left\{(x,y) \in S | \dot{\Delta} = \frac{\partial \Delta}{\partial y} \times \kappa(x,y) = 0, \kappa(x,y) \neq 0\right\} \qquad (6)$$

Of particular importance are $n-2$ dimensional connected components in $\Xi$ which can be divided into semi-saddle $\Xi^{sa}$ and semi-focus $\Xi^{fo}$. The dynamics in the vicinity of $\Xi^{sa}$ are shown in Figure 1 where the trajectories "curve" towards $\Gamma^s$ i.e. $\ddot{\Delta}(x,y) > 0$ whereas for $\Xi^{fo}$, they curve away. Obviously, only $\Xi^{sa}$ can exist on the quasi stability boundary.

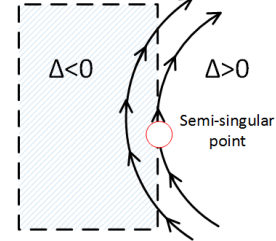

*Figure 1 Dynamics Near Semi-Saddle*

The important second category of points are called pseudo equilibrium points which are EPs of system (5) but not of system (1) as defined below.
$$\psi = \{x,y \in S | f(x,y) \neq 0, \kappa(x,y) = 0\} \qquad (7)$$

Of particular importance are $n-2$ dimensional connected components of transverse pseudo EPs $\psi^{tr}$. These points have $n-2$ dimensional center manifold which is the connected component of pseudo EPs itself and two non-zero eigen values with the associated eigen vectors transversally intersecting with $S$ and thus the name transverse. Depending on the sign of those eigen values, the points can be characterized as source (both positive) $\psi^{trso}$, sink (both negative) $\psi^{trsi}$ or saddle (one positive one negative) $\psi^{trsa}$. Connected component of $\psi^{trsa}$ are crucial for characterizing the quasi-stability boundary. The dynamics in its vicinity are shown in Figure 2 where the arrows point in the direction of the flow.

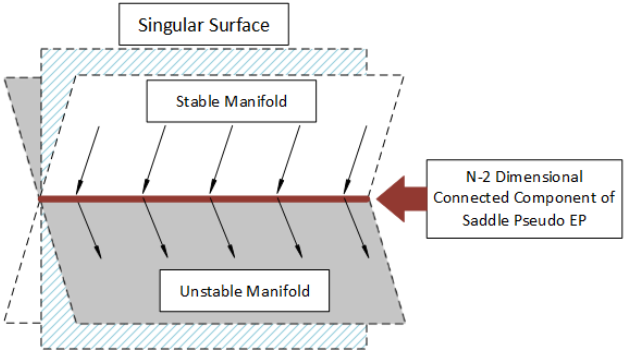

*Figure 2 Dynamics near Transverse Saddle Pseudo EP*

Besides that, the stability boundary also contains some $n-1$ dimensional components of $S$ that repel the trajectories in their vicinity (on the $\Gamma^s$ side) towards $(x^s, y^s)$. Under reasonable assumptions [15], the quasi-stability boundary of the transformed system (5) and consequently the original DAE system in (1) is comprised of –

1. $n-1$ dimensional components of $S$
2. $n-1$ dimensional set of points in $\Gamma^s$ on which the trajectories intersect $S$ at one of the following,
    a. $n-2$ dimensional component of $\Xi^{sa}$
    b. $n-2$ dimensional component of $\psi^{trsa}$
3. stable manifolds of type-1 UEP and periodic orbits

## III. CCT Sensitivity Derivation

### A. Overview

The overall goal is to derive the expressions for sensitivity of CCT of any given fault to variations in any parameter $p$ evaluated at a base value $p^*$ i.e. $\left.\frac{\partial CCT}{\partial p}\right|_{p=p^*}$ for a generalized parameter varying DAE system of the form,
$$\dot{x} = f(x, y, p) \qquad (8)$$
$$0 = g(x, y, p)$$

The term base critical trajectory will refer to the trajectory (fault-on + post-fault) obtained for the system with $p = p^*$ when the fault under study is cleared at its CCT. We will be using the superscript "$*$" to represent the values of various variables along the base critical trajectory. For example, if a generic fault-clearing time is denoted by $t^{cl}$, its value for the base critical trajectory will be denoted by $t^{cl^*}$.

Since TSA is only concerned with the post-fault system's stability, we will only focus on that system's stability boundary. By definition, for $t^{cl} = CCT$, the state variable value at $t^{cl}$ lies on one of the four types of components on the stability boundary as listed at the end of Section II. B. On varying $p$, the stability boundary will change and so will the fault-on trajectory. Therefore, for the new fault-on trajectory to intersect the new stability boundary, the fault clearing time will have to adjusted where the amount of adjustment required per unit change in $p$ is the sensitivity of CCT. The overall process is as follows -

1. Find the base critical trajectory (using TDS).
2. Find the sensitivity of the relevant component of the stability boundary [16].
3. Find the sensitivity of state variable values at the time of fault clearing to $p$ and $t^{cl}$.
4. Equate the above two to get CCT sensitivity.

The main challenges in the above procedure are as follows –
   i. The closed form expression for the stability boundary is usually not available and only local approximations can be made around critical points. For example, the equation of stable manifold of type-1 UEP is locally approximated near the UEP by a hyperplane normal to the unstable eigen vector.
   ii. The local approximation listed above is usually not given around the point where the fault trajectory intersects the stability boundary (exit point) but is available at some other point along the post-fault trajectory.

Challenge i) is straightforward to deal with since we are only calculating first order sensitivities and therefore can replace the first step with calculating the sensitivity of local approximation to the stability boundary. Challenge ii) requires modifying the third step of the approach to evaluate the sensitivity of state variable values at that point on the base critical trajectory where the local approximation of the stability boundary is available.

### B. Dealing with Discontinuity in y

In a typical TSA study, the system passes through at least three distinct system conditions viz. pre-fault, fault-on and post-fault with appropriate subscripts chosen to differentiate the active system conditions. As discussed before, these discrete changes result in the $y$ value jumping between different constraint surfaces denoted by $\Gamma_{pre}, \Gamma_{fault}$ and $\Gamma_{post}$ which introduces complications. An easy way out is to construct an extended state system to keep track of the image of $y$ on the active surface (which governs the current dynamics) as well as the surface it can potentially jump to. Let the corresponding images at any time $t$ be denoted by $y_{pre}(t), y_{fault}(t)$ and $y_{post}(t)$. As an example, when analyzing the fault on dynamics, we need to track not only $x(t)$ and $y_{fault}(t)$ but also $y_{post}(t)$ which physically represents the value of $y$ immediately following the clearing of fault at any time $t$. This is necessary since the stability of the post-fault system and therefore the overall system depends on where $(x, y_{post})$ is w.r.t post-fault SEP's stability region. The state equation of this *extended state fault-on system* can easily be written as,
$$\dot{x} = f_{fault}(x, y_{fault}, p) \qquad (9)$$
$$0 = g_{fault}(x, y_{fault}, p)$$
$$0 = g_{post}(x, y_{post}, p)$$

Notice how $y_{post}$ dynamics during fault are coupled with $x$ whose own dynamics are governed by $y_{fault}$.

### C. Sensitivity of the State Value at Fault Clearing

Let $x^0, y^0_{pre}$ and $y^0_{fault}$ denote the state values (and its image on $\Gamma_{fault}$) at $t = 0$. Usually, the system is assumed to starts from pre-fault system's SEP $(x^s_{pre}, y^s_{pre})$ which gives,
$$f_{pre}(x^0, y^0_{pre}, p) = 0 \qquad (10)$$
$$g_{pre}(x^0, y^0_{pre}, p) = 0$$
$$g_{fault}(x^0, y^0_{fault}, p) = 0$$

Differentiating the first two equations above and evaluating at the starting point of the base critical trajectory $(x^{0^*}, y^{0^*}_{pre})$,

$$\left.\frac{\Delta x^0}{\Delta p}\right|_{p^*} = A_1^{(n\times 1)} = \left[\frac{\partial f_{pre}}{\partial x^0} - \frac{\partial f_{pre}}{\partial y_{pre}^0} \times \left[\frac{\partial g_{pre}}{\partial y_{pre}^0}\right]^- \times \frac{\partial g_{pre}}{\partial x^0}\right]^- \quad (11)$$
$$\times \left(\frac{\partial f_{pre}}{\partial y_{pre}^0} \times \left[\frac{\partial g_{pre}}{\partial y_{pre}^0}\right]^- \times \frac{\partial g_{pre}}{\partial p}\right.$$
$$\left.\left. - \frac{\partial f_{pre}}{\partial p}\right)\right|_{x^{0*},y_{pre}^{0*},p^*}$$

Next, we compute the sensitivity $x^{cl}$ which is the $x$ value at fault clearing time. We know that $(x^0, y_{fault}^0)$ cannot be a singular point on $\Gamma_{fault}$ otherwise the fault-on trajectory would not exist. Therefore, by implicit function theorem [13], $y_{fault}^0$ can locally be written as a function of $x^0$ and consequently, $x^{cl} = \varphi_{fault}^x(x^0, t^{cl}, p)$ where $\varphi_{fault}^x$ represents the flow of $x$ coordinate for the fault-on system. The sensitivity of $x^{cl}$ is evaluated at the base system's CCT $t^{cl*}$ as follows,

$$\left.\frac{\Delta x^{cl}}{\Delta p}\right|_{p^*} = B_1 \times \left.\frac{\Delta x^0}{\Delta p}\right|_{p^*} + B_2 \times \left.\frac{\Delta t^{cl}}{\Delta p}\right|_{p^*} + B_3 \quad (12)$$

Where,
$B_1^{(n\times n)} = \left.\frac{\partial \varphi_{fault}^x}{\partial x^0}\right|_{x^{0*},t^{cl*},p^*}, B_2^{(n\times 1)} = \left.\frac{\partial \varphi_{fault}^x}{\partial t}\right|_{x^{0*},t^{cl*},p^*} = f_{fault}(x, y_{fault}, p)|_{x^{cl*},y_{fault}^{cl*},p^*}, B_3^{(n\times 1)} = \left.\frac{\partial \varphi_{fault}^x}{\partial p}\right|_{x^{0*},t^{cl*},p^*}$

$B_1$ and $B_3$ are solutions to the variational equations for the fault-on base system ($p = p^*$) ([17]) given below.

$$\frac{\partial \varphi_{fault}^x}{\partial \alpha} = \frac{\partial f_{fault}}{\partial x} \times \frac{\partial \varphi_{fault}^x}{\partial \alpha} + \frac{\partial f_{fault}}{\partial y_{fault}} \times \frac{\partial \varphi_{fault}^{y_{fault}}}{\partial \alpha} + \frac{\partial f_{fault}}{\partial \alpha} \quad (13)$$
$$0 = \frac{\partial g_{fault}}{\partial x} \times \frac{\partial \varphi_{fault}^x}{\partial \alpha} + \frac{\partial g_{fault}}{\partial y_{fault}} \times \frac{\partial \varphi_{fault}^{y_{fault}}}{\partial \alpha} + \frac{\partial g_{fault}}{\partial \alpha}$$

For $B_1$, with $\alpha = x^0$ and $\frac{\partial \varphi_{fault}^x}{\partial \alpha}(0) = I^{(n\times n)}$ while for $B_3$, $\alpha = p$ and $\frac{\partial \varphi_{fault}^x}{\partial \alpha}(0) = 0^{(n\times 1)}$. $\frac{\partial \varphi_{fault}^{y_{fault}}}{\partial \alpha}(0)$ is obtained from the second equation above. Substituting (11) in (12) we get,

$$\boxed{\left.\frac{\Delta x^{cl}}{\Delta p}\right|_{p^*} = B_1 \times A_1 + B_2 \times \left.\frac{\Delta t^{cl}}{\Delta p}\right|_{p^*} + B_3} \quad (14)$$

### D. Singularity Immediately Following Fault Clearing

In this instability scenario, the base critical fault-on trajectory becomes unstable by directly intersecting the singular surface. Therefore, CCT also represents the time it takes for the sustained fault trajectory to reach singularity/voltage collapse. There is a greater value to deriving this sensitivity for each fault regardless of the phenomenon for instability for the base critical trajectory as this number will give an insight into what control parameters are effective in *pushing away* the singular surface thereby reducing the likelihood of voltage collapse.

Now, the fault-on system is assumed to not have any singularities within the region of interest. Therefore, the phenomenon being studied is one where the fault-on trajectory intersects $S_{post}$ immediately on clearing the fault i.e. $(x^{cl}, y_{post}^{cl}) \in S_{post}$. Here $y_{post}^{cl}$ denotes the $y$ value right after clearing the fault at $t = t^{cl}$. Therefore, $(x^{cl}, y_{post}^{cl})$ satisfies,

$$\left.\frac{\partial \Delta_{post}(x,y,p)}{\partial x}\right|_{x^{cl*},y_{post}^{cl*},p^*} \times \left.\frac{\Delta x^{cl}}{\Delta p}\right|_{p^*} + \left.\frac{\partial \Delta_{post}(x,y,p)}{\partial y}\right|_{x^{cl*},y_{post}^{cl*},p^*} \times \quad (15)$$
$$\left.\frac{\Delta y_{post}^{cl}}{\Delta p}\right|_{p^*} + \left.\frac{\partial \Delta_{post}(x,y,p)}{\partial p}\right|_{x^{cl*},y_{post}^{cl*},p^*} = 0$$

$$\left.\frac{\partial g_{post}(x,y,p)}{\partial x}\right|_{x^{cl*},y_{post}^{cl*},p^*} \times \left.\frac{\Delta x^{cl}}{\Delta p}\right|_{p^*} + \left.\frac{\partial g_{post}(x,y,p)}{\partial y}\right|_{x^{cl*},y_{post}^{cl*},p^*} \times$$
$$\left.\frac{\Delta y_{post}^{cl}}{\Delta p}\right|_{p^*} + \left.\frac{\partial g_{post}(x,y,p)}{\partial p}\right|_{x^{cl*},y_{post}^{cl*},p^*} = 0$$

Since $\left.\frac{\partial g_{post}(x,y,p)}{\partial y}\right|_{x^{cl*},y_{post}^{cl*},p^*}$ is singular, let there be a left eigen vector $v^{sing*T}$ corresponding to the 0 eigen value. Pre-multiplying the second equation above gets rid of the second term yielding,

$$v^{sing*T} \times C_1 \times \left.\frac{\Delta x^{cl}}{\Delta p}\right|_{p^*} + v^{sing*T} \times C_2 = 0 \quad (16)$$

Where $C_1^{(m\times n)} = \left.\frac{\partial g_{post}(x,y,p)}{\partial x}\right|_{x^{cl*},y_{post}^{cl*},p^*}$ and $C_2^{(m\times 1)} = \left.\frac{\partial g_{post}(x,y,p)}{\partial p}\right|_{x^{cl*},y_{post}^{cl*},p^*}$. Substituting (14) in (16) gives the final expression for CCT sensitivity,

$$\boxed{\left.\frac{\Delta t^{cl}}{\Delta p}\right|_{p^*} = -\frac{v^{sing*T} \times (C_2 + C_1 \times (B_1 \times A_1 + B_3))}{v^{sing*T} \times C_1 \times B_2}} \quad (17)$$

### E. Singularity in Post-Fault Trajectory

In this section, the instability phenomenon involves the base critical post-fault trajectory eventually intersecting $S_{post}$ as opposed to immediately intersecting as studied in the previous case. From discussions in Section II. B. , the point of intersection of the fault-on trajectory with $S_{post}$ can either be a semi-saddle point $\Xi_{post}^{sa}$ or a transverse saddle pseudo EP $\psi_{post}^{trsa}$.

As discussed before, the post-fault trajectory ceases to exist on $S_{post}$ and therefore the end point of the post-fault trajectory lies either on an $n-2$ dimensional connected component of $\Xi_{post}^{sa}$ or $\psi_{post}^{trsa}$. Since the local characterization of the relevant component of the stability boundary can be derived around the critical element itself, the first step is to estimate the sensitivity of a generalized end point of the post-fault trajectory $(x^{end}, y_{post}^{end})$ which is clearly a function of its starting point $(x^{cl}, y_{post}^{cl})$, the time spent along the post-fault trajectory $t^{end}$ and obviously $p$. i.e. $x^{end} = \varphi_{post}^x((x^{cl}, y_{post}^{cl}), t^{end}, p)$ and $y_{post}^{end} = \varphi_{post}^y((x^{cl}, y_{post}^{cl}), t^{end}, p)$. Similar to the argument used previously, $(x^{cl}, y_{post}^{cl})$ has to be a regular point and thus $y_{post}^{cl}$ can locally be written as a function of $x^{cl}$. Computing and evaluating the sensitivity at the base critical trajectory we get,

$$\left.\frac{\Delta x^{end}}{\Delta p}\right|_{p^*} = D_1 \times \left.\frac{\Delta x^{cl}}{\Delta p}\right|_{p^*} + D_2 \times \left.\frac{\Delta t^{end}}{\Delta p}\right|_{p^*} + D_3 \quad (18)$$

Where,
$D_1^{(n\times n)} = \left.\frac{\partial \varphi_{post}^x}{\partial x^{cl}}\right|_{x^{cl*},t^{end*},p^*}, D_2^{(n\times 1)} = \left.\frac{\partial \varphi_{post}^x}{\partial t}\right|_{x^{cl*},t^{end*},p^*} = f_{post}(x, y_{post}, p)|_{x^{end*},y_{post}^{end*},p^*}, D_3^{(n\times 1)} = \left.\frac{\partial \varphi_{post}^x}{\partial p}\right|_{x^{cl*},t^{end*},p^*}$

$D_1$ and $D_3$ are solutions to the variational equations for the post-fault base system ($p = p^*$) as done previously for equation (13). Combining (18) with (14) yields,

$$\boxed{\begin{aligned}\left.\frac{\Delta x^{end}}{\Delta p}\right|_{p^*} &= D_1 \times B_2 \times \left.\frac{\Delta t^{cl}}{\Delta p}\right|_{p^*} + D_2 \times \left.\frac{\Delta t^{end}}{\Delta p}\right|_{p^*} \\ &\quad + (D_3 + D_1 \times (B_1 \times A_1 + B_3))\end{aligned}} \quad (19)$$

Now, both $\Xi_{post}^{sa}$ and $\psi_{post}^{trsa}$ lie on $S_{post}$. Furthermore, these are $n-2$ dimensional and therefore locally defined by $n+m+2$ equality constraints of the form, $\{(x,y) \in R^{m+n} | \Delta_{post}(x,y,p) = 0, g_{post}(x,y,p) = 0, \lambda_{post}(x,y,p) = 0\}$ where $\Delta_{post}(x,y,p)$ is a generalized scalar function. Thus, $(\frac{\Delta x^{end}}{\Delta p}, \frac{\Delta y_{post}^{end}}{\Delta p})$ is characterized by,

$$\begin{bmatrix} E_1, E_2 \\ F_1, F_2 \\ G_1, G_2 \end{bmatrix} \times \begin{bmatrix} \frac{\Delta x^{end}}{\Delta p}\Big|_{p^*} \\ \frac{\Delta y_{post}^{end}}{\Delta p}\Big|_{p^*} \end{bmatrix} = -\begin{bmatrix} E_3 \\ F_3 \\ G_3 \end{bmatrix} \quad (20)$$

Where,
$E_1^{(1\times n)} = \frac{\partial \Delta_{post}(x,y,p)}{\partial x}\Big|_{x^{end^*}, y_{post}^{end^*}, p^*}$, $E_2^{(1\times m)} = \frac{\partial \Delta_{post}(x,y,p)}{\partial y}\Big|_{x^{end^*}, y_{post}^{end^*}, p^*}$,

$E_3^{(1\times 1)} = \frac{\partial \Delta_{post}(x,y,p)}{\partial p}\Big|_{x^{end^*}, y_{post}^{end^*}, p^*}$, $F_1^{(m\times n)} = \frac{\partial g_{post}(x,y,p)}{\partial x}\Big|_{x^{end^*}, y_{post}^{end^*}, p^*}$

$F_2^{(m\times m)} = \frac{\partial g_{post}(x,y,p)}{\partial y}\Big|_{x^{end^*}, y_{post}^{end^*}, p^*}$, $F_3^{(m\times 1)} = \frac{\partial g_{post}(x,y,p)}{\partial p}\Big|_{x^{end^*}, y_{post}^{end^*}, p^*}$

$G_1^{(1\times n)} = \frac{\partial \lambda_{post}(x,y,p)}{\partial x}\Big|_{x^{end^*}, y_{post}^{end^*}, p^*}$, $G_2^{(1\times m)} = \frac{\partial \lambda_{post}(x,y,p)}{\partial y}\Big|_{x^{end^*}, y_{post}^{end^*}, p^*}$

$G_3^{(1\times 1)} = \frac{\partial \lambda_{post}(x,y,p)}{\partial p}\Big|_{x^{end^*}, y_{post}^{end^*}, p^*}$

Finally, combining (19) with (20), we get the expression for CCT sensitivity for base critical post fault trajectory intersecting the $S_{post}$ at either $\Xi_{post}^{sa}$ or $\psi_{post}^{trsa}$.

$$\begin{bmatrix} E_1 D_1 B_2 & E_1 D_2 & E_2 \\ F_1 D_1 B_2 & F_1 D_2 & F_2 \\ G_1 D_1 B_2 & G_1 D_2 & G_2 \end{bmatrix} \times \begin{bmatrix} \frac{\Delta t^{cl}}{\Delta p}\Big|_{p^*}, \frac{\Delta t^{end}}{\Delta p}\Big|_{p^*}, \frac{\Delta y_{post}^{end}}{\Delta p}\Big|_{p^*} \end{bmatrix}^T$$
$$= \begin{bmatrix} -E_3 - E_1(D_3 + D_1(B_1 A_1 + B_3)) \\ -F_3 - F_1(D_3 + D_1(B_1 A_1 + B_3)) \\ -G_3 - G_1(D_3 + D_1(B_1 A_1 + B_3)) \end{bmatrix} \quad (21)$$

For semi-saddle points ($\Xi_{post}^{sa}$), in the above equations, $\lambda_{post}(x,y,p) = \frac{\partial \Delta_{post}}{\partial y} \times \kappa_{post}$. As for transverse saddle pseudo EPs $\psi_{post}^{trsa}$, $(x^{end}, y_{post}^{end}) \in \{(x,y) \in R^{m+n} | \Delta_{post}(x,y,p) = 0, g_{post}(x,y,p) = 0, \kappa_{post}(x,y,p) = 0\}$ which is a set characterized by $2m+1$ equality constraints i.e. having even an $n-1-m$ dimensional component. However, since only $n-2$ dimensional $\psi_{post}^{trsa}$ exist on the quasi-stability boundary [15], $rank\left(\left[\left[\frac{\partial g_{post}}{\partial x}, \frac{\partial g_{post}}{\partial y}\right]; \left[\frac{\partial \Delta_{post}}{\partial x}, \frac{\partial \Delta_{post}}{\partial y}\right]; \left[\frac{\partial \kappa_{post}}{\partial x}, \frac{\partial \kappa_{post}}{\partial y}\right]\right]\right) = m+2$. Furthermore, by assumptions stated in [15] which are generally met, $\left[\frac{\partial g_{post}}{\partial x}, \frac{\partial g_{post}}{\partial y}\right]$ is full ranked in $\Gamma_{post}^s \cup S_{post}$. Therefore, there exists a scalar function $\kappa_{1_{post}}$ which is an element of vector function $\kappa_{post}$ s.t. the rows of $\left[\left[\frac{\partial g_{post}}{\partial x}, \frac{\partial g_{post}}{\partial y}\right]; \left[\frac{\partial \Delta_{post}}{\partial x}, \frac{\partial \Delta_{post}}{\partial y}\right]; \left[\frac{\partial \kappa_{1_{post}}}{\partial x}, \frac{\partial \kappa_{1_{post}}}{\partial y}\right]\right]$ span those of the previous matrix which results in $\lambda_{post}(x,y,p) = \kappa_{1_{post}}(x,y,p)$. Now, in order to ensure numerical stability of the equation (21), $\kappa_{1_{post}}$ is chosen as the scalar component function of $\kappa_{post}$ that maximizes the smallest singular value of the above matrix.

### F. Post-Fault Trajectory Converging to a Type-1 UEP

In this instability scenario, the critical post-fault trajectory eventually converges to a type-1 UEP with singularity not playing any role. This usually manifests in the form of loss of synchronism of generator(s). Here, only the key points of the derivation will be shown with further details in [11].

Let $(x^{cu}, y_{post}^{cu})$ be the CUEP for the given fault which by definition is a regular point. The stable manifold of $(x^{cu}, y_{post}^{cu})$ can locally be approximated by the following hyperplane,

$$(v^{cu}(p))^T \times (x^{end} - x^{cu}(p)) = 0 \quad (22)$$

Where $v^{cu}$ is the only unstable eigen vector of the reduced state matrix, $\left[\frac{\partial f_{post}}{\partial x} - \frac{\partial f_{post}}{\partial y_{post}} \times \left[\frac{\partial g_{post}}{\partial y_{post}}\right]^- \times \frac{\partial g_{post}}{\partial x}\right]\Big|_{x^{cu}, y_{post}^{cu}, p}$.

Now, we know that $\lim_{t^{end} \to \infty}(x^{end}, y_{post}^{end}) \to (x^{cu}, y_{post}^{cu})$. Therefore, differentiating (22), combining with (19) and evaluating as $(x^{end^*}, y_{post}^{end^*}) \to (x^{cu^*}, y_{post}^{cu^*})$, we get the final expression for CCT sensitivity.

$$\boxed{\frac{\Delta t^{cl}}{\Delta p}\Big|_{p^*} = -\frac{v^{cu^{*T}} \times (H_1 - (D_3 + D_1 \times (B_1 \times A_1 + B_3)))}{v^{cu^{*T}} \times D_1 \times B_2}} \quad (23)$$

Where, $H_1^{(n\times n)} = \left[\frac{\partial f_{post}}{\partial x} - \frac{\partial f_{post}}{\partial y_{post}} \times \left[\frac{\partial g_{post}}{\partial y_{post}}\right]^- \times \frac{\partial g_{post}}{\partial x}\right]^- \times$
$\left(\frac{\partial f_{post}}{\partial y_{post}} \times \left[\frac{\partial g_{post}}{\partial y_{post}}\right]^- \times \frac{\partial g_{post}}{\partial p} - \frac{\partial f_{post}}{\partial p}\right)\Big|_{x^{cu^*}, y_{post}^{cu^*}, p^*}$

## IV. NOTES ON NUMERICAL IMPLEMENTATION AND COMPUTATIONAL ASPECTS

When trying to implement the overall process, there could be a few numerical challenges which will be discussed here. When the mechanism of instability of the base critical trajectory is the traditional one like loss of synchronism, it is nearly impossible to ensure that the fault is cleared precisely on the stable manifold of the type-1 CUEP. Furthermore, it is highly likely for the same trajectory to eventually hit the singular surface. Therefore, in order to correctly identify the instability mechanism so as to use the appropriate CCT sensitivity expression, it is advisable to first check whether the unstable trajectory passes close to a UEP by evaluating $\|f_{post}\|$.

When the mode of instability is singularity in the post-fault phase, the simulation stops converging on reaching $S_{post}$. Depending on the precision of the underlying solver, the simulation might end considerably far from $S_{post}$ which could introduce errors when evaluating CCT sensitivity expression in (21). This can be resolved by extrapolating $\Delta_{post}$ to detect 0 crossing. Similarly, when the fault-on trajectory directly intersects $S_{post}$, simulating the extended state fault-on system given by (9) will cause non-convergence at $S_{post}$. However, this adds on $m$ more equations which slows down the overall simulation. A way out could be to only compute $y_{post}$ at some points along the fault-on trajectory till $\Delta_{post}$ changes sign. Thereafter, the time and state values $(x^{cl}, y_{post}^{cl})$ at zero crossing of $\Delta_{post}$ can be found by interpolating the fault-on trajectory.

It is usually advisable that $(x^{end^*}, y_{post}^{end^*})$ be sufficiently close to the appropriate critical element on the stability boundary in order to get an accurate estimate of the CCT sensitivity. However, a high precision comes at a cost of a greater number of iterations. In our experience, a high precision in base critical trajectory is usually only required if the given $p^*$ value is associated with a topological change to the relevant stability boundary i.e. CCT is nearly non-differentiable w.r.t $p$.

When using this approach for large scale systems, the computation of trajectory sensitivities is a major portion of the overall computation. Luckily, this step can be made extremely efficient by using parallel programming and sparsity techniques as shown in [18] for practical systems. Furthermore, our derivations are based on characterizations of the stability boundary which hold true in general [14]. As a justification, Direct methods for TSA [7] which are based on the same characterizations have shown great promise in terms of reliability for real-time TSA of large-scale systems [19].

## V. RESULTS

In this section, we validate our derived expressions by comparing them against CCT values obtained through repetitive TDS under varying parameter values. Two of the standard test systems ([14]) which have popularly been used for studying the stability theory for DAE systems will be used to demonstrate the validity of our derived expressions. These systems are low dimensional and therefore easy to visualize.

### A. Case 1: Example 7-5 [14]

A parameterized version of system in Example 7-5 of [14] is given along with other relevant functions.

$$f_{post}(x,y,p) = \begin{bmatrix} y^2 + py - 2x_1 + x_2 + 1 \\ -x_2 \end{bmatrix} \quad (24)$$

$$g_{post}(x,y,p) = x_1 y - py - x_2 + y^3$$

$$\Delta_{post}(x,y,p) = 3y^2 - p + x_1$$

$$\kappa_{post}(x,y,p) = 2 - y \times (p - x_1 + 1) - x_2$$

$$\frac{\partial \Delta_{post}}{\partial y} \times \kappa_{post} = 12y + 6x_1 y^2 - (6x_2 y + 6py^2 + 6y^2)$$

Let us study the variation of $p^*$ from -0.4 to 0.4. This system is particularly interesting due to the presence of both a semi-saddle and a transverse saddle pseudo EP on the stability boundary for the range of $p$ values under study. In order to validate the expression in (21), we define the fault-on dynamics as $f_{fault}(x,y,p) = [x_2, -1]^T$, $g_{fault}(x,y,p) = g_{post}$ such that it take the system towards the semi-saddle point.

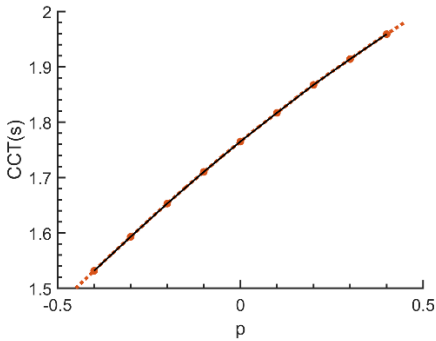

Figure 3 Example 7-5 CCT vs p

The CCT vs $p$ curve is plotted in Figure 3. The bold orange colored points correspond to distinct $p^*$ values with the color orange denoting the mode of instability which is the post-fault trajectory intersecting a semi-saddle point. At each point, a dotted straight line is drawn with slope equal to the CCT sensitivity value computed using (21). This serves as the local estimate of CCT vs $p$ curve. It can be seen that the local estimate is in fact tangential to the actual curve which validates the expression in (21).

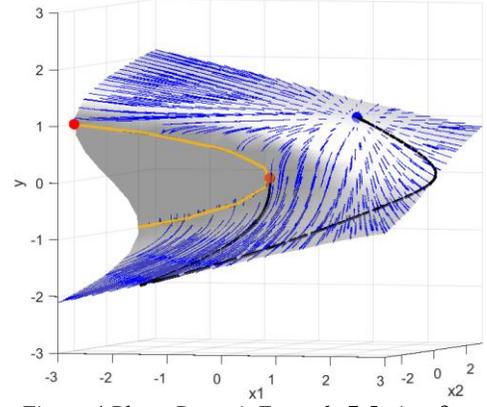

Figure 4 Phase Portrait Example 7-5 $p^* = 0$

To get visual insights, we plot the phase portrait of this system with its base critical trajectory for $p^* = 0$ (black curve) in Figure 4. The SEP of interest is marked in blue, the relevant semi-saddle point at (0, 0, 0) is marked orange and a transverse saddle pseudo EP is marked in red at (-3, -2, 1). The singular surface is traced using a yellow line. The base critical trajectory can be seen intersecting the singular surface tangentially.

### B. Case 2: One Machine One Bus System

Next, we validate the remaining expressions using a one bus one machine model with bus voltage angle taken as a reference.

$$f_{post}(x,y,p) = \begin{bmatrix} x_2 + \dfrac{P_m - \dfrac{E \times y}{X}\sin(x_1)}{D_l} \\ \dfrac{\left(P_m - \dfrac{E \times y}{X}\sin(x_1) - D_g \times x_2\right)}{M} \end{bmatrix} \quad (25)$$

$$g_{post}(x,y,p) = \frac{E \times y}{X}\cos(x_1) - \frac{y^2}{X} - Q_l$$

Here, $x_1$ is the deviation of generator rotor angle from bus phase angle, $x_2$ is the generator angular speed deviation and $y$ is the bus voltage magnitude. The parameters $p$ comprise of the generator inertia constant $M$, mechanical power input to the generator $P_m$, generator damping $D_g$, internal emf of the generator $E$, the reactive power load at the bus $Q_l$, the load damping factor $D_l$ and $X$, the total series impedance (internal impedance of generator plus transmission line impedance).

The fault being studied is a 3 phase to ground fault on the bus i.e. $g_{fault}(x,y,p) = y = 0$ which is cleared without changing the network topology (pre-fault and post-fault systems are same). Let us study the variation of $P_m$ on CCT. Let $p^* = [X = 0.5, P_m = 0.3: 0.5, E = 1, M = 1, D_l = 1, D_g = 1, Q_l = 0.1]$.

In Figure 5, CCT is plotted vs $P_m$ as done previously. As expected, CCT reduces with increasing generator loading due to the SEP moving closer to the stability boundary. Here, the points corresponding to various values of $p^* = P_m^*$ are colored red or green depending on whether the mechanism of instability of the base critical trajectory is loss of synchronism or base critical post-fault trajectory intersecting a transverse saddle pseudo EP respectively. This means that the mechanism for instability changes as $P_m$ goes beyond 0.4 and therefore we the appropriate CCT sensitivity expressions are used i.e. (21) for red points and (23) for green. The local estimates obtained from those expressions are shown using dotted lines as done previously. Clearly, the estimates are tangent to the real curve which validates both (21) and (23).

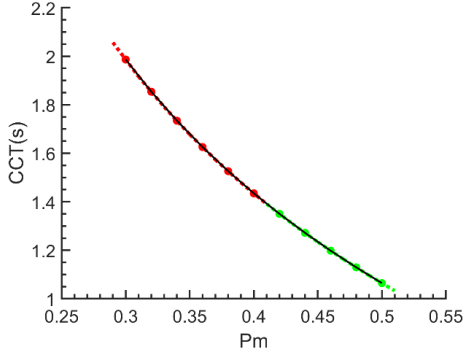

Figure 5 CCT vs $P_m$

To get a closer look into this transition in the instability mechanism due to variations in $P_m$, we plot the phase portrait along with the base critical trajectories (black curve) for $P_m$ =0.3 and 0.5 in Figure 6 and Figure 7 respectively. The relevant portion of the singular surface in these cases can be seen as a one-dimensional component (yellow line) forming the nose of the constraint surface. Near/on the singular surface is a pseudo EP (red) and a UEP (green).

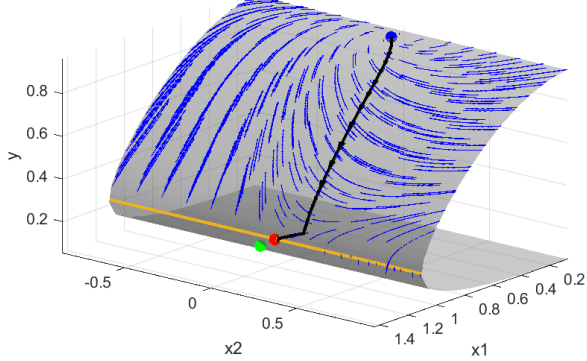

Figure 6 Phase Portrait Single Machine $P_m = 0.3$

From Figure 6, for $P_m = 0.3$, the dynamics in the vicinity of the pseudo EP clearly show that it is a transverse saddle type. The UEP near it does not lie on the stability boundary since it is beyond the singular surface (under the yellow line).

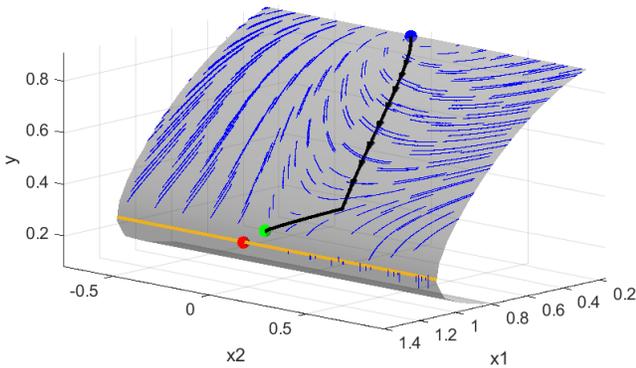

Figure 7 Phase Portrait Single Machine $P_m = 0.5$

As $P_m$ increases to 0.5, the previously irrelevant UEP (green) crosses the singular surface and now becomes a type – 1 UEP lying on the stability boundary. The transverse saddle type pseudo EP (green) now becomes a source type as it repels the trajectories in its vicinity. Therefore, instability now happens through loss of synchronism as opposed to singularity which can also be seen from the base critical trajectory. A very important observation that can be made by comparing Figure 6 to Figure 7 is that what was earlier the stable manifold of transverse saddle pseudo EP now becomes the stable manifold of the type-1 UEP. Therefore, the relevant portion of the stability boundary is actually the same manifold even after a sudden change in the instability mechanism. Now, this manifold changes smoothly with $p$ which is its natural behavior and this explains why the CCT vs $p$ curve remains smooth even at the point of transition of instability mechanism at $P_m = 0.4$.

Next, we modify the load model to be frequency dependent resulting in $g_{post}(x,y,p) = \frac{E \times y}{X}\cos(x_1) - \frac{y^2}{X} - Q_l \times (1 + x_2)$ and study the effects of changes in inertia $M$ with increasing renewable generation on stability. Let, $p^* = [X = 0.5, P_m = 0.5, E = 1, M = 0.1:0.4, D_l = 1, D_g = 1, Q_l = 0.1]$. CCT is plotted against $M$ in Figure 8 along with the sensitivity estimates given by yellow (fault-on trajectory directly intersecting the singular surface) and green (loss of synchronism) dotted lines obtained using (17) and (23) respectively. The estimates are clearly tangential to the actual curve which validates the expressions in (17) and (23).

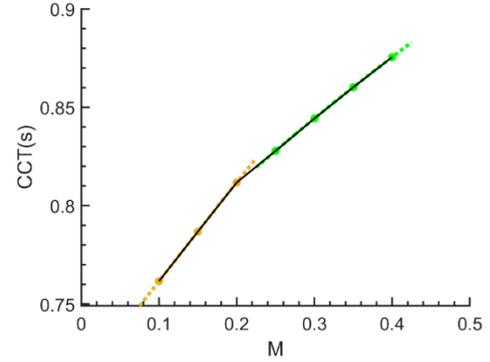

Figure 8 CCT vs M (Frequency Dependent Load)

Unlike the previous case, the CCT vs $p = M$ curve is not differentiable at the point of transition of instability mechanism between $M = 0.2$-$0.25$. To explain this, we plot the phase portrait for $M = 0.2$ in Figure 9.

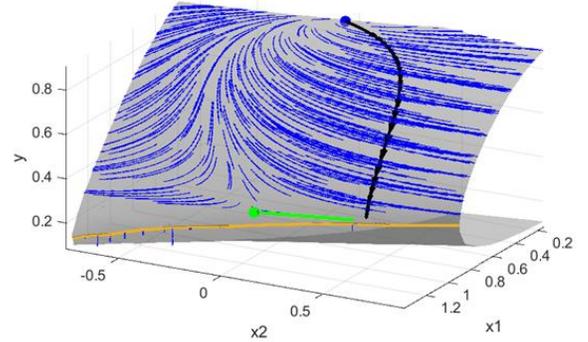

Figure 9 Phase Portrait Single Machine Frequency Dependent Load M=0.2

Here, the relevant portion of the local stable manifold of type 1 UEP (green) is marked with a bold green line which intersects the singular surface (yellow curve) transversally [13]. From (25), it can be seen that $M$ has no effect on $g_{post}$ and consequently the singular surface while it does influence the stable manifold of the type-1 UEP meaning that they have very different behaviors to changes in $M$. As $M$ varies, the relevant portion of the stability boundary switches from a surface not sensitive to $M$ to one that is sensitive resulting in the CCT vs $M$ curve being non-differentiable at the point of transition. As discussed before, in such cases, the CCT sensitivity expressions could suffer from ill-conditioning problems when evaluating at $p^*$ close to the transition point thus requiring high precision.

## VI. Conclusions and Future Work

In this paper, given a critical fault-on and post-fault trajectory, we derived expressions for sensitivity of CCT to parameter variations for DAE systems. Besides the traditional instability mechanism of loss of synchronism of generator(s), DAE type models also exhibit a phenomenon where the system trajectory reaches a region in state space marked by singularity of algebraic constraints which is closely related to voltage collapse. This is particularly relevant to the TSA of systems operating in weak conditions with some unmodeled dynamics resulting in such singularities influencing the size of the stability region. Due to multiple possible mechanisms of instability, the appropriate CCT sensitivity expression is derived for each. The derived expressions were shown to be valid when compared with the computationally intensive repeated TDS. It was also observed that when studying the effect of parameters that unequally impact neighboring components of the stability boundary, there can be situations where the instability mechanism changes under parameter variations resulting in a non-differentiable CCT vs parameter curve. These scenarios were found to require high precision due to the ill conditioning of the CCT sensitivity expressions when evaluated at parameter values close to the transition points.

A potential application of this work could be for identifying effective controls for enhancing CCT for some critical faults suffering from voltage collapse. Another application could be for TSA of networks having uncertainties in operating conditions such as high renewable penetration systems. These will be explored in future.